\documentclass[preprint,12pt]{elsarticlem}

\biboptions{square,comma,numbers}

\newcounter{bla}

\begin{document}

\begin{frontmatter}

\title{QCDMAPT\underline{~~}F: Fortran version of QCDMAPT package}

\author[a]{A.V.~Nesterenko\corref{author}}
\author[b]{C.~Simolo}

\cortext[author]{Corresponding author.\\
\textit{E-mail address:} nesterav@theor.jinr.ru}

\address[a]{BLTPh, Joint Institute for Nuclear Research,
Dubna, 141980, Russian Federation}

\address[b]{ISAC--CNR, I--40129, Bologna, Italy}

\begin{abstract}
The QCDMAPT program package facilitates computations in the framework of
dispersive approach to Quantum Chromodynamics. The QCDMAPT\underline{~~}F
version of this package enables one to perform such computations with
Fortran, whereas the previous version was developed for use with Maple
system. The QCDMAPT\underline{~~}F package possesses the same basic
features as its previous version. Namely, it embodies the calculated
explicit expressions for relevant spectral functions up to the four--loop
level and the subroutines for necessary integrals.
\end{abstract}

\end{frontmatter}

\noindent
{\bf NEW VERSION PROGRAM SUMMARY}

\vskip2.5mm

\begin{small}
\noindent
{\em Manuscript Title:}
QCDMAPT\underline{~~}F: Fortran version of QCDMAPT package    \\
{\em Authors:} A.V.~Nesterenko and C.~Simolo                  \\
{\em Program Title:} QCDMAPT\underline{~~}F                   \\
{\em Journal Reference:} Comput.\ Phys.\ Comm.\ \textbf{182}
  (2011) 2303--2304                                           \\
  %Leave blank, supplied by Elsevier.
{\em Catalogue identifier:} AEGP\underline{~~}v2\underline{~~}0 \\
  %Leave blank, supplied by Elsevier.
%{\em Licensing provisions:} none                              \\
  %enter "none" if CPC non-profit use license is sufficient.
{\em Programming language:} Fortran~77 and higher             \\
{\em Computer:} Any which supports Fortran~77                 \\
  %Computer(s) for which program has been designed.
{\em Operating system:} Any which supports Fortran~77         \\
  %Operating system(s) for which program has been designed.
%{\em RAM:} bytes                                             \\
  %RAM in bytes required to execute program with typical data.
%{\em Number of processors used:}                             \\
  %If more than one processor.
%{\em Supplementary material:}                                \\
  % Fill in if necessary, otherwise leave out.
{\em Keywords:} Nonperturbative QCD; Dispersion relations     \\
  % Please give some freely chosen keywords that we can use
  % in a cumulative keyword index.
{\em Classification:} 11.1, 11.5, 11.6                        \\
  %Classify using CPC Program Library Subject Index, see (
  % http://cpc.cs.qub.ac.uk/subjectIndex/SUBJECT_index.html)
  %e.g. 4.4 Feynman diagrams, 5 Computer Algebra.
{\em External routines/libraries:} MATHLIB routine RADAPT~(D102) from
CERNLIB Program Library~\cite{MATHLIB}. \\
  % Fill in if necessary, otherwise leave out.
%{\em Subprograms used:}                                      \\
  %Fill in if necessary, otherwise leave out.
{\em Catalogue identifier of previous version:}
AEGP\underline{~~}v1\underline{~~}0 \\
  %Only required for a New Version summary, otherwise leave out.
{\em Journal reference of previous version:}
Comput.\ Phys.\ Comm.\ \textbf{181} (2010) 1769--1775                  \\
  %Only required for a New Version summary, otherwise leave out.
{\em Does the new version supersede the previous version?:} No \\
  %Only required for a New Version summary, otherwise leave out.
{\em Nature of problem:} A central object of the dispersive (or
``analytic'') approach to Quantum Chromodynamics~\cite{APT,MAPT} is the
so--called spectral function, which can be calculated by making use of the
strong running coupling. At the one--loop level the latter has a quite
simple form and the relevant spectral function can easily be calculated.
However, at the higher loop levels the strong running coupling has a
rather cumbersome structure. Here, the explicit calculation of
corresponding spectral functions represents a somewhat complicated task
(see Sect.~3 and App.~B of Ref.~\cite{QCDMAPT}), whereas their numerical
evaluation requires a lot of computational resources and essentially slows
down the overall computation process. \\
  %Describe the nature of the problem here.
{\em Solution method:} The developed package includes the calculated
explicit expressions for relevant spectral functions up to the four--loop
level and the subroutines for necessary integrals. \\
  %Describe the method solution here.
{\em Reasons for the new version:} The previous version of the package
(Ref.~\cite{QCDMAPT}) was developed for use with Maple system. The new
version is developed for Fortran programming language. \\
  %Only required for a New Version summary, otherwise leave out.
{\em Summary of revisions:} The QCDMAPT\underline{~~}F package consists of
the main program (QCDMAPT\underline{~~}F.f) and two samples of the file
containing the values of input parameters (QCDMAPT\underline{~~}F.i1 and
QCDMAPT\underline{~~}F.i2). The main program includes the definitions of
relevant spectral functions and subroutines for necessary integrals. The
main program also provides an example of computation of the values of
(M)APT spacelike/timelike expansion functions for the specified set of
input parameters and (as an option) generates the output data files with
values of these functions over the given kinematic intervals. \\
  %Only required for a New Version summary, otherwise leave out.
%{\em Restrictions:} \\
  %Describe any restrictions on the complexity of the problem here.
%{\em Unusual features:} \\
  %Describe any unusual features of the program/problem here.
{\em Additional comments:} For the proper functioning of
QCDMAPT\underline{~~}F package, the ``MATHLIB'' CERNLIB
library~\cite{MATHLIB} has to be installed. \\
  %Provide any additional comments here.
{\em Running time:} The running time of the main program with sample set
of input parameters specified in the file QCDMAPT\underline{~~}F.i2 is
about a minute (depends on~CPU). \\
  %Give an indication of the typical running time here.

\end{small}


\begin{thebibliography}{0}

\bibitem{MATHLIB}
Subroutine D102 of the ``MATHLIB'' CERNLIB library, URL addresses: \\
http://cernlib.web.cern.ch/cernlib/mathlib.html \\
http://wwwasdoc.web.cern.ch/wwwasdoc/shortwrupsdir/d102/top.html
  %%CITATION = NONE;%%

\bibitem{APT}
D.V.~Shirkov and I.L.~Solovtsov, Phys.\ Rev.\ Lett.\ \textbf{79} (1997) 1209; \\
  %%CITATION = PRLTA,79,1209;%%
K.A.~Milton and I.L.~Solovtsov, Phys.\ Rev.\ D \textbf{55} (1997) 5295; \\
  %%CITATION = PHRVA,D55,5295;%%
K.A.~Milton and I.L.~Solovtsov, Phys.\ Rev.\ D \textbf{59} (1999) 107701; \\
  %%CITATION = PHRVA,D59,107701;%%
I.L.~Solovtsov and D.V.~Shirkov, Theor.\ Math.\ Phys.\ \textbf{120} (1999) 1220; \\
  %%CITATION = TMFZA,120,482;%%
D.V.~Shirkov and I.L.~Solovtsov, Theor.\ Math.\ Phys.\ \textbf{150} (2007) 132.
  %%CITATION = TMPHA,150,132;%%

\bibitem{MAPT}
A.V.~Nesterenko, Phys.\ Rev.\ D \textbf{62} (2000) 094028; \\
  %%CITATION = PHRVA,D62,094028;%%
A.V.~Nesterenko, Phys.\ Rev.\ D \textbf{64} (2001) 116009; \\
  %%CITATION = PHRVA,D64,116009;%%
A.V.~Nesterenko, Int.\ J.\ Mod.\ Phys.\ A \textbf{18} (2003) 5475; \\
  %%CITATION = IMPAE,A18,5475;%%
A.V.~Nesterenko, Nucl.\ Phys.\ B (Proc.\ Suppl.) \textbf{133} (2004) 59; \\
  %%CITATION = NUPHZ,133,59;%%
A.V.~Nesterenko and J.~Papavassiliou, J.\ Phys.\ G \textbf{32} (2006) 1025; \\
  %%CITATION = JPHGB,G32,1025;%%
A.V.~Nesterenko, Nucl.\ Phys.\ B (Proc.\ Suppl.) \textbf{186} (2009) 207; \\
  %%CITATION = NUPHZ,186,207;%%
A.V.~Nesterenko, arXiv:1106.4006~[hep-ph].
  %%CITATION = ARXIV:1106.4006;%%

\bibitem{QCDMAPT}
A.V.~Nesterenko and C.~Simolo, Comput.\ Phys.\ Commun.\ \textbf{181} (2010) 1769.
  %%CITATION = CPHCB,181,1769;%%

\end{thebibliography}
\end{document}